\documentstyle[amssymb,psfig]{article}

\newcommand{\resection}[1]{\setcounter{equation}{0}\section{#1}}
\newcommand{\appsection}{\addtocounter{section}{1} \setcounter{equation}{0}
                         \section*{Appendix \Alph{section}}}

\def\de {\mbox{d}}
\def\be{\begin{equation}}
\def\ee{\end{equation}}
\def\bd{\begin{displaystyle}}
\def\ed{\end{displaystyle}}
\def\ba{\begin{array}}

\def\ea{\end{array}}
\def\EQ{\begin{equation}}
\def\EN{\end{equation}}
\def\bea{\begin{eqnarray}}
\def\eea{\end{eqnarray}}
\def\beano{\begin{eqnarray*}}
\def\eeano{\end{eqnarray*}}
\def\to{\rightarrow}

\def\th{\theta}
\def\nn{\nonumber}
\begin{document}
\setcounter{page}{0}
\renewcommand{\thefootnote}{\arabic{footnote}}
\newpage
\setcounter{page}{0}

\begin{titlepage}

\begin{flushright}
ISAS/EP/2/97\\
\end{flushright}
\vspace*{0.5cm}

\begin{center}
{\bf 
\begin{Large}
{\bf
Form Factors of Exponential Operators and \\
Exact Wave Function Renormalization Constant\\
in the Bullough--Dodd Model \\}
\end{Large}
}
\vspace*{1.5cm}{\large C. Acerbi\footnote{e--mail: acerbi@sissa.it} }
         \\[.3cm]
         {\it Scuola Internazionale Superiore di Studi Avanzati (SISSA)}\\
         {\it and}\\
 {\it Istituto Nazionale di Fisica Nucleare}\\
          {\it 34014 Trieste, Italia}
\end{center}
\vspace*{0.7cm}
\begin{abstract}
\noindent
We compute  the form factors of exponential operators
$e^{kg\varphi(x)}$ in the two--dimensional integrable 
Bullough--Dodd model ($a_2^{(2)}$ Affine Toda Field Theory). 
These form factors are selected 
among the solutions of general nonderivative scalar operators 
by their asymptotic cluster property. 
Through analitical continuation to complex values of the coupling 
constant these solutions permit  to  compute  
the form factors of  scaling relevant primary fields  
in the lightest--breather sector of integrable 
$\phi_{1,2}$ and $\phi_{1,5}$ deformations of conformal minimal models.
We also obtain the exact wave--function renormalization constant
$Z(g)$ of the model and the properly normalized form factors 
of the operators $\varphi(x)$ and $:\!\varphi^2(x)\!:$ .
\end{abstract}
\vfill
\end{titlepage}
\setcounter{footnote}{0}

\resection{Introduction}
In  recent years much important progress has 
been achieved in 
the study of two dimensional Quantum Field Theory 
and related Statistical Mechanical systems.
The solution of Conformal Field Theories \cite{BPZ,ISZ} has not only 
allowed the full characterizaton
of fixed points in the renormalization group describing 
universality classes 
of critical models, but  it has also  
provided the possibility of describing 
the renormalization group flow away from criticality 
by means of  relevant 
deformations of Conformal Minimal Models \cite{ZamC}. 
In particular, Zamolodchikov showed that in 
some  interesting cases \cite{Zam} ---
falling in the classes of  $\phi_{1,2}$, $\phi_{1,3}$ and 
$\phi_{2,1}$   deformations --- 
infinitely many integrals of 
motion survive the deformation and the system is suitably 
described by an integrable relativistic scattering theory.
Bootstrap techniques relying on the integrability of the 
model provide then a powerful tool for obtaining the exact 
$S$--matrix of the system which turns
out to be elastic and factorizable \cite{ZamZam,GM}.
A systematic description of the $S$-matrices 
for all the integrable deformed minimal conformal models 
has been given \cite{Sm0,Sm1,BLC,Sm2} in terms of specific 
reductions of the 
two only existing two--dimensional single--boson integrable  models,
namely the sinh--Gordon and the Bullough--Dodd (BD) model (the Affine Toda 
Field Theories $a^{(1)}_1$ and
$a^{(2)}_{2}$) in their complex coupling constant versions
which are also referred to as the 
sine--Gordon model and the Zhiber--Mihailov--Shabat (ZMS) model
respectively \cite{BD,ZMS,MOP}.

It is widely  believed that  
the knowledge of the scattering data 
amounts to an exhaustive  solution 
of a quantum field theory and that, 
in principle, one should be able to recover from them 
the operator content of the theory as well as to compute
the correlation  functions of local operators.
In order to carry out this task,
the so--called form factors approach
has been developed 
and successfully employed in many 
important cases \cite{BKW,Smi,YL,YZ,CM,DM,AMV,DS,BN}.
The strategy  of this technique relies in   
the reconstruction of correlation functions 
by means of a spectral sum\footnote{For notational
convenience we assume here the spectrum to consist 
of a single particle $A$ and parameterize the momenta 
in terms of the rapidity variable $\th$.} 
\be \label{spectral}
 \langle 0| \Phi_1(x) \, \Phi_2(0)|0 \rangle = 
 \sum_n \,\frac{1}{n!} \int\, 
 \frac{d\th_1}{2\pi} \cdots \frac{d\th_n}{2\pi}\, 
 F_n^{\Phi_1}(\th_1,\ldots,\th_n)
  \, \left[F_n^{\Phi_2}(\th_1,\ldots,\th_n)\right]^* 
  \, e^{-m|x|\sum_{i} \cosh\th_i}  \: ,
\ee
on all the intermediate $n$--particle states of 
a scattering theory involving 
on--shell amplitudes of local operators 
(the so--called form factors)
\[ 
F_n^{\Phi} (\th_1,\ldots,\th_n) = 
\langle 0| \Phi(0)| A(\th_1) \ldots A(\th_n) \rangle \,.
\]
Form factors in turn  
can be exactly obtained 
in two dimensional bootstrap systems
as solutions 
of a system of functional equations which entail the 
correct analiticity and monodromy 
properties dictated by the $S$--matrix.
The space of solutions of the system is supposed to give a faithful 
representation of the operatorial content of the theory \cite{CM2},
but the correct identification of the form factors 
of a specific operator within this space of solutions 
is in general a nontrivial problem.
A useful criterion for this identification was given in \cite{DM}
where it was proved that for a scaling operator $\Phi$ of 
scaling dimensions $2\Delta_\Phi$ the
form factors divergence   for large values of the 
rapidities is bounded by 
\[
\lim_{|\th_i |\to \infty} F_n^\Phi(\th_1,\ldots,\th_n) \lesssim  
e^{\Delta_\Phi |\th_i |}   \, \, .
\]
More recently, it has been observed in \cite{DSC} that the form factors 
of relevant ($\Delta_\Phi<1$) scaling operators satisfy 
a simple factorization property given by the so--called 
``cluster equations''
\bea \label{cluster}
\lim_{\Delta\to \infty} \,F_n^{\Phi}(\th_1+\Delta,\ldots,
\th_m+\Delta,\th_{m+1},\ldots,\th_{n}) 
 = \frac{1}{F_0^{\Phi}} \, \, F_m^{\Phi}(\th_1,\ldots,\th_m) 
 \, F_{n-m}^{\Phi}
 (\th_{m+1},\ldots,\th_n)  \,, 
\eea 
($\forall m=1,\ldots,n-1$), 
which hold unless some internal symmetry of the theory
makes some of the form factors vanish.
This property had already been noticed to be satisfied 
in the solutions of some  specific models \cite{Sm0,YL,KM,MS}
and is believed to be a distinguishing property 
of exponential operators in Lagrangian theories.
In particular, in ref.~\cite{KM}  a  family of cluster solutions
was found  in the sinh--Gordon model and further identified 
with the form factors of  exponential operators. 
These solutions were then used to compute the form factors 
of primary operators \cite{K} in a class of 
$\phi_{1,3}$--deformed minimal models in which 
the boson of the original Lagrangian theory 
is still present after reduction.

In the present paper we analyze the form factors 
of scalar operators in  
the Bullough--Dodd model and  focus in particular 
on  possible solutions of cluster equations 
(\ref{cluster}). In view  of the above discussion,
these solutions become particularly interesting,
not only as candidate solutions for the identification 
of exponential operators  of the model,
but also because in the complex 
coupling constant version (ZMS model) 
one should be able to identify among them 
the scaling relevant  primary fields 
in the  reductions which 
describe specific deformations of minimal models.

The paper will be organized as follows.
In Section 2 we review some general features of the BD model and 
its intepretation as a Complex Liouville Theory that allows us 
to map the exponential operators into the scaling primary fields 
of the $\phi_{1,2}$, $\phi_{2,1}$ and $\phi_{1,5}$ deformed minimal models.
In Section 3  we analyze the general solution 
of  form factors  equations for  non--derivative scalar 
operators of the BD model, 
exhibiting the first multiparticle form factors 
and giving a full characterization of the dimensionality of the space
of solutions.
In Section 4 we study a one parameter family of cluster
solutions which is shown to correspond to the form factors of exponential 
operators $e^{kg\varphi(x)}$ and we  determine the exact formula, 
eq. (\ref{ansatz}), which gives the  dependence on $k$ and $g$ 
of these solutions.
In Section 5 we check the validity of the correct interpretation
of cluster solutions by comparing
our results with all the known cases of form factors 
of scaling primary operators computed 
in $\phi_{1,2}$ and $\phi_{1,5}$ 
deformations of minimal models.
In Section 6 we make use of the 
knowledge of the form factors of exponential operators 
to exactly compute 
the wave function renormalization constant  of the BD model 
and the form factors of the fields $\varphi(x)$ and $:\!\varphi^2(x)\!:$
correctly normalized.
Finally, we draw our conclusions in Section 7.

\resection{The Bullough--Dodd model and its interpretation  
           as a Complex Liouville Theory}
The so--called Bullough--Dodd  (BD) model \cite{BD} is a
           two--dimensional 
integrable 
Lagrangian QFT, namely the $a_2^{(2)}$  Affine 
Toda Field Theory \cite{MOP},
defined by the Lagrangian density
\be \label{L}
{\cal L} =  \frac{1}{2} \left( \partial_\mu\varphi\right)^2
- \frac{m_0^2}{6 \, g^2} \,
 \left( 2 \,:\!e^{g\varphi}\!: + :\!e^{-2 g \varphi}\!:\right) \, ,
\ee
where  the exponentials are normal ordered.
This model  is the only  2D integrable QFT involving a single 
bosonic field which exhibits  the 
$\varphi^3$ property (i.e. the elementary particle  appears as a bound
state of itself). The spectrum of the theory consists of a single 
bosonic massive particle $A$ of mass $m$ and 
the exact $S$--matrix of the model 
factorises into two--particle amplitudes given by the following 
function of the relative rapidity variable $\th$ \cite{AFZ}
\[
S(\th) = 
\:f_{\frac{2}{3}}(\th) 
\:f_{\frac{B-2}{3}}(\th) 
\:f_{-\frac{B}{3}}(\th)\, ,
\]
where we have used the building block function
\[
 f_{x}(\th) \equiv \frac{\tanh\frac{1}{2}(\th + i\pi\,x)}
                   {\tanh\frac{1}{2}(\th - i\pi\,x)} \, .
\]
The renormalized coupling constant $B$ is given by
\[ 
B(g)= \frac{g^2/2\pi}{1+g^2/4\pi}\, ,
\]
and ranges from $0$ to $2$  for real values of $g$. 
The $S$--matrix exhibits a weak--strong coupling duality
under the transformation $g\longleftrightarrow 4\pi/g$ 
or equivalently $B\longleftrightarrow 2-B$.
For later use we also define  the following
duality--invariant function of the coupling constant
\be \label{c}
 c = \cos \frac{(B+2)\, \pi}{3}.
\ee
The $S$--matrix has a simple pole at $\th= 2\pi i /3$ 
corresponding to the bound state represented by the particle $A$
itself.
The on--shell  three--point coupling constant is given by
\[
\Gamma^2 = -i \lim_{\theta\rightarrow 2 i \pi/3}
(\theta - \frac{2 \pi i}{3})\,
S(\theta) = 2 \,\sqrt{3}\, \frac{(c+1)(1+2c)}{(c-1)(1-2c)}\:,
\]
and vanishes both at the free field limiting  values $B=0,2$ and
at the self--dual point $B=1$.

For imaginary values of the coupling constant $g$  (i.e. $B<0$),
the BD model --- which is then  referred to as 
the Zhiber--Mihailov--Shabat (ZMS) model \cite{ZMS} --- 
permits the description of  $\phi_{1,2}$ and $\phi_{2,1}$ 
deformations of conformal minimal models.
Indeed,  starting from the observation 
that the ZMS has a non--unitary 
$S$--matrix related to the Izergin--Korepin $R$--matrix,
Smirnov exploited the 
quantum group $SL(2)_q$ invariance of the  $S$--matrix 
in order to recover unitarity in specific reductions
of the model.
The $S$--matrices of the above--mentioned deformed 
minimal models were in this way  obtained from   RSOS restrictions of the 
Izergin--Korepin  $R$--matrix at specific values of the coupling constant 
at which  $q$ is a  root of unity \cite{Sm2}.
More recently the possibility of studying some relevant
$\phi_{1,5}$  deformations of 
specific non--unitary minimal models 
has been considered as well which relies again on 
quantum group reductions of the ZMS model \cite{Tac}.

Some insight can be obtained  if one considers the ZMS model by
interpreting one of the exponential operators in the Lagrangian 
as a deformation of a Complex Liouville Theory (CLT) \cite{DF}.  
The other exponential then plays  the role of a screening operator.
If we require the CLT to describe the minimal model ${\cal M}_{r,s}$ 
with central charge
\[
C = 1- \frac{6\,(r-s)^2}{r\,s} \, , \:\:\:\:\:\: s>r \:\:\:\:
 \mbox{\it relative primes} \, ,
\] 
and  primary fields $\phi_{m,n}$ of 
conformal dimensions  
\[
\Delta_{m,n} = \frac{(n\,r-m\,s)^2-(r-s)^2}{4\,r\,s} \:\: \:\:\:\:\: m=1,\ldots r-1 ; \:\: n = 1,\ldots s-1 \,,
\]
the above interpretation leads to a four--fold choice: in fact,  
after choosing one of the two exponentials in eq. (\ref{L}) as the 
screening operator, one can still choose two possible values
of $g$ as  a function of $r$ and $s$ in order to correctly 
set  its conformal dimensions to be $\Delta=1$.
In the Complex Liouville Theory, the primary operators  
will be given by the following exponential operators\footnote{In the
following we will always consider  normal ordered 
exponential operators omitting the notation $:\!e^{\alpha \varphi(x)}\!:$.}
\be \label{map}
\phi_{m,n} = e^{k_{m,n} g  \varphi} \:\: \:\:\:\:\: 
m=1,\ldots r-1 ; \:\: n = 1,\ldots s-1 ;
\ee
with the identification 
\be \label{symkac}
\phi_{m,n} \equiv \phi_{r-m,s-n} \, ,
\ee
which entails the correct symmetry of the Kac Table of
the model.
In eq. (\ref{map}), 
the dependence of $k_{m,n}$ and $B(g)$ from the integers 
$r$, $s$, $m$, $n$ depends
on the choice made for the screening operator, 
as summarized in Table 1 and  the deforming 
exponential can be easily shown to correspond to one of the primary  
fields $\phi_{1,2}$,   $\phi_{2,1}$, $\phi_{1,5}$ or  $\phi_{5,1}$.
However, one can easily check that, while the primary field $\phi_{1,2}$ 
is relevant in any minimal model,
the field $\phi_{5,1}$ is on the contrary always irrelevant 
and therefore does not yield renormalizable deformations.
As for the fields $\phi_{2,1}$ and $\phi_{1,5}$, they can be shown to be relevant 
only in disjoint sets of models:
the field $\phi_{2,1}$ is relevant for the class of minimal models 
${\cal M}_{r,s}$ with $s<2r$ 
which includes all the unitary cases ${\cal M}_{r,r+1}$, 
while  $\phi_{1,5}$ is relevant for the complementary class of non--unitary models $s>2r$.
Notice that in order to  decide  whether the deformation is relevant or not
it is sufficient to   require that  the coupling constant $g$  be
imaginary, namely that $B<0$ (see Table 1).

The spectrum of the reduced ZMS model in general 
consists of kinks together with a
cascade of their possible bound states \cite{Sm2}. 
This  spectrum does not always contain the 
original BD boson: while this particle is always present in the 
$\phi_{1,2}$ deformations (where it appears as the lightest 
breather of two fundamental kinks),  it is on the contrary 
never present in the spectrum of 
$\phi_{2,1}$ deformations. 
In the relevant $\phi_{1,5}$ deformations 
the presence of the BD boson  depends on the specific model (see \cite{Tac}).

We will not enter in further details on the reductions of the ZMS model
which can be found in the original literature \cite{Sm2,Tac}.
The information collected in this chapter is all we need 
for establishing the correct mapping between exponential operators of the BD model 
and primary fields of the reduced models.

\resection{Form Factor Equations in the BD model}
We now turn to  the problem of determining the on--shell matrix elements (form factors)
of a  local operator ${\Phi(x)}$ in the BD model.
In the framework of two--dimensional integrable QFT, 
this problem is reduced to the problem 
of studying a set of coupled functional equations \cite{BKW,Smi} namely the Watson 
monodromy equations  
\be\label{Watson}
\ba{l}
F_n (\th_1,\ldots,\th_i,\th_{i+1},\ldots,\th_n) = S(\th_i-\th_{i+1}) \, 
F_n (\th_1,\ldots,\th_{i+1},\th_{i},\ldots,\th_n) \, ,\\ 
F_n (\th_1+ 2\pi i,\th_2\ldots,\th_n) = F_n (\th_2\ldots,\th_n,\th_1)
\, ,
\ea
\ee
as well as the recursive residue equations 
on  annihilation poles (kinematical residue equations) 
\be \label{kinF}
\lim_{\th^\prime- \th} \,
F_{n+2} (\th^\prime + i\pi,\th ,\th_1,\ldots,\th_n) = i\,
\left( 1 - \prod_{i=1}^{n} S(\th-\th_i)\right)\, F_{n} (\th_1,\ldots,\th_n) \, ,
\ee
and bound state poles (dynamical residue equations) 
\be \label{dynF}
\lim_{\alpha \to \frac{2\pi i}{3}} \left(\alpha - \frac{2\pi i}{3}\right) \, 
F_{n+2} (\th+ \alpha/2,\th-\alpha/2 ,\th_1,\ldots,\th_n) = i\, \Gamma 
F_{n+1} (\th,\th_1,\ldots,\th_n) \, .
\ee
The most general solution to the 
monodromy equations (\ref{Watson}) can be written in the following 
form \cite{BKW}
\[
F_n^{\Phi}(\th_1,\ldots,\th_n)= R^{\Phi}(\th_1,\ldots,\th_n)\: \prod_{i<j} 
F^{\min}(\th_{i}-\th_{j})
 \:,
\]
where $R^{\Phi}(\th_1,\ldots,\th_n)$ 
is any symmetric $2\pi i$--periodic function 
in the variables $\th_i$
and the ``minimal''  two--particle form factor $F^{min}(\th)$ 
is given by the following function 
\be  \label{Fmin}
   F^{min}(\th) = {\cal N}(B) \frac{ g_{0}(\th)\:g_{\frac{2}{3}}(\th) }
                          { g_{\frac{2-B}{3}}(\th)
\:g_{\frac{B}{3}}(\th) } \, ,
\ee
where  $g_{\alpha}(\th)$ is defined by 
\[
g_{\alpha}(\th)=
\exp\left[2\int_0^{\infty}\frac{\de t}{t}
\frac{\cosh\left((\alpha-1/2)t\right)}{\cosh (t/2) \sinh t}
\sin^2((i\pi-\th)\,t/2\pi)\right]\, .
\]
In eq. (\ref{Fmin}),  ${\cal N}(B)$ is the following
normalization constant 
\be \label{N}
{\cal N}(B) = \exp 
\left[ -4 \,\int \, \frac{dt}{t} \,
 \frac{\sinh (t/2)\, \cosh (t/6)}{\sinh^2 t}
\, \left( \cosh (t/3) - \cosh ((B-1)\, t/3) \right) \right] \, ,
\ee
chosen such that $F^{min}(\infty)= 1$.
For real values of the coupling constant, namely for $B\in (0,2)$, 
$F^{min}(\th)$ has neither poles nor zeros in the physical strip ${\rm Im} \th\in (0,\pi)$,
since the same property is shared by  $g_{\alpha}(\th)$ when $\alpha \in (0,1)$.
The analitical continuation of   $F^{min}(\th)$ for imaginary values
of the  coupling constant $g$ ($B <0$) developes poles in $\th$ which can be 
explicitly exhibited  by using the following functional relations 
\bea
& & g_{1+\alpha}(\th) = g_{-\alpha}(\th) \, , \nonumber\\
& & g_{\alpha}(\th) \, g_{-\alpha}(\th) = {\cal P}_{\alpha}(\th) 
\equiv \frac{\cos\pi\alpha - \cosh\th}
{2 \cos^2\frac{\pi\alpha}{2}} \, ,\nonumber
\eea
satisfied by the functions $g_\alpha(\th)$.

Notice that  we have not mentioned yet
the dependence of the form factors $F^\Phi_n$
on the operator $\Phi(x)$. Indeed, in the system of equations (\ref{Watson}), (\ref{kinF}) 
and (\ref{dynF}) this dependence is not explicit and further physical
requirements  are necessary to identify 
in the space of solutions the form factors of a specific operator.

\subsection{General solutions for scalar non--derivative operators}
In this work we are mainly concerned with the analysis of form
factors of scalar operators which are local nonderivative functions
of the field $\varphi(x)$. This infinite dimensional operatorial space 
can be spanned for instance 
by  the basis of polynomials in $\varphi (x)$ or
by the basis of exponentials $e^{\alpha\varphi(x)}$.
A suitable parameterization of the form factors for this class of
operators is the following
\be \label{Fn}
F_n^{\Phi}(\th_1,\ldots,\th_n)= H_n^{\Phi} \, Q_n^{\Phi}(x_1,\ldots,x_n)\: \prod_{i<j} 
\frac{F^{min}(\th_{i}-\th_{j})}
{(x_i + x_j)(x_i^2  + x_i x_j +x_j^2)} \:,
\ee
where   $x_i = e^{\th_i}$.
The pole structure expected to reflect the correct analiticity properties 
 is explicitly shown in the denominator of (\ref{Fn}), where
annihilation and bound state simple poles are present at relative rapidities
$\th_{ij}= i\pi$ and $\th_{ij}= 2 \pi i/3$, respectively.
$Q_n^\Phi$ is a  homogeneous symmetrical 
polynomial in the variables $x_i$ whose   total degree  
is determined by  Lorentz invariance  to be
$d_n = \frac{3 \, n \, (n-1)}{2}$.
The constants $H_n^{\Phi}$  are conveniently chosen to be 
\be  \label{H}
 H_n^{\Phi} = t \,\mu^n(B) \:,
\ee 
in order to obtain a simplified  version of recursive equations 
on the polynomials $Q_n^{\Phi}$.
In eq. (\ref{H}), $t$ is  a free  parameter which will have 
an important role in the discussion of cluster solutions whereas  
\[
  \mu(B) = \frac{\sqrt{3}\,\, \Gamma(B) }{F^{min}(\frac{2 \pi i}{3})} .
\]
With the  above choice of $H_n^{\Phi}$,  
the dynamical recursive equations (\ref{dynF}) read 
\be \label{dyn}
 Q_{n}(\omega x, \omega^{-1} x, x_1,\ldots,x_{n-2}) = 
- \, x^3 
 D_{n-2}(x|x_1,\ldots,x_{n-2}) \, Q_{n-1}(x,x_1,\ldots,x_{n-2}) \, ,
\ee
where $\omega= e^{i \pi/3}$ and  the polynomial $D_n$ is given by 
\be \label{D}
D_n(x|x_1,\ldots,x_n)= \sum_{k_1,k_2,k_3=0}^{n} \, x^{3n- k_1-k_2-k_3} \,
\sigma_{k_1}^{(n)}\, \sigma_{k_2}^{(n)}\,
 \sigma_{k_3}^{(n)}\,\cos \left((k_2-k_3)(B+2)\pi/3
\right) \:.
\ee
The last expression is written in 
the usual basis of symmetrical polynomials $\sigma_k^{(n)}$
which are defined by the generating function
\be \nonumber
 \sum_{k=0}^n \, x^{n-k} \,\sigma_k^{(n)} = \prod_{i=1}^{n} \, (x+ x_i) \, .
\ee
In  expression (\ref{D}) we can get rid of the trigonometrical dependence 
on the coupling constant $B$ by exploiting the following recursive relation
\[
\cos ((n+1)\alpha) = 2\, \cos (n \alpha) \, \cos \alpha - \cos ((n-1)
\alpha) \:, 
\]
which allows us to express cosines of multiple angles as polynomials
of $\cos\alpha$. In this way   
we can  cast the dependence  of eq. (\ref{dyn}) on the coupling constant into 
a {\em rational} dependence on the variable $c$ defined in eq. (\ref{c}).

The kinematical residue equations on annihilation poles (\ref{kinF})
can be written as  
\be \label{kin}
Q_{n} (-x,x,x_1,\ldots,x_{n-2}) = (-)^n \,K\, x^3 \, 
U_{n-2}(x|x_1,\ldots,x_{n-2}) \, Q_{n-2} (x_1,\ldots,x_{n-2}) \:, 
\ee
with 
\bea
U_n(x|x_1,\ldots,x_n) &=& 2 \sum_{k_1,\ldots,k_6=0}^n (-)^{k_2+k_3+k_5} \,
x^{6n-(k_1+ \cdots +k_6)} \,
\sigma_{k_1}^{(n)} \cdots \sigma_{k_6}^{(n)} \cdot \\ \nonumber & & \cdot
\sin\left( \left( 2\, (k_2 + k_4 -k_1 - k_3) + 
B\,(k_3 + k_6 - k_4 - k_5) \right) \pi/3 \right) \:,
\eea
and 
\[
K = \frac{(2c-1)}{4\,\sqrt{3} \,(1+c)(2c+1)} \:.
\]
Before solving the system of recursive equations,
let us  derive some important properties on the space of solutions 
from a direct analysis of the equations (\ref{dyn}) and (\ref{kin}).
\begin{description}
\item[A --] 
It is easy to prove that in the space of symmetrical polynomials 
of degree $d_n = \frac{3 \, n \, (n-1)}{2}$, the only polynomials 
which have zeros both at $x_i/x_j = e^{2\pi i/3}$ and at $x_i/x_j= -1$ are
given by  
\bea
 {\cal K}^{(n)}(\{x_i\})&=&  \prod_{1\leq i<j \leq n} (x_i+x_j)
     (x_i^2  +x_i\,x_j+ x_j^2) \nonumber
\\ &=& \det \left|\sigma_{2\, j - i }^{(n)} \right|_{1\leq
i,j \leq  n-1} \,\,\,\, \det \left|\sigma_{3 [j/2]- i + 1 + (-)^{j+1} }^{(n)} \right|_{1\leq
i,j \leq 2\, n-2}  \,\, . \nonumber
\eea
up to a multiplicative constant.
This is therefore the only possible kernel for the whole 
system of recursive equations. Hence, after fixing all the polynomials
$Q_i$ for $i=1,\ldots n-1$, the most general solution $Q_n$ of the 
system of equations (\ref{dyn}) and (\ref{kin}) will be then given by
\be \label{affine}
Q_{n} = Q_{n}^* + \lambda_{n} \, {\cal K}^{(n)}(\{x_i\}) \:,
\ee
where $Q_{n}^*$ is a specific solution and $\lambda_n$ is a free parameter.
The space of solutions will be organized correspondingly, namely 
every operator will be identified by a succession of parameters 
$\lambda_i$, $i=1,\ldots \infty$
and the general solution for a $n$--particle form factor will be described 
by an $n$--dimensional vector space of solutions $Q_n$ 
spanned by the parameters $\lambda_1,\ldots,\lambda_n$. 
\item[B --]
The partial degree of the general polynomial $Q_n$ with respect 
to any of the variables $x_i$ is exactly $d_n^{(i)} =  3 (n-1)$.
This can be easily shown by induction 
observing that $Q_1$ must be a constant for Lorentz invariance
and making use of equations  (\ref{dyn}),  (\ref{kin}) and  (\ref{affine}).
This implies in particular that the form factors of this class 
of scalar operators  of  the theory  have bounded asymptotic 
behavior for large values of the rapidities,
\[
\lim_{\Lambda \rightarrow \infty} 
       F^{\Phi}_{n} 
       (\th_1+\Lambda,\ldots,\th_k+\Lambda, \th_{k+1},
       \ldots, \th_{n}) < 
\infty \:\:\:\:\:\:\:\:\:\:\:\forall k = 1,\ldots,n-1.
\] 
This observation enables us to look for cluster solutions of form factors
equations within this general class of solutions (see eq. (\ref{cluster})).
\end{description}
We now turn to the actual computation of the 
first multiparticle   general solutions 
to the system of recursive equations (\ref{dyn}) and (\ref{kin}). 
The most direct way of computing these solutions consists in 
parameterizing any polynomial $Q_n$ 
as the most general polynomial of degree $d_n = \frac{3 \, n \, (n-1)}{2}$ in the basis 
of symmetrical polynomials $\sigma_k^{(n)}$ and to impose 
on the coefficients of the expansion the costraints coming from the recursive equations.
We report here the result of the first general multiparticle
form factors in the space of scalar non--derivative operators.
Lorentz invariance requires  $Q_1$ to be a constant 
\[
Q_1= \lambda_1\:,
\]
hence in order not to have two different overall normalization 
constants we can set for the time being $t=1$ in eq. (\ref{H}).
The next most general solutions are  given by
\[
Q_2(x_1,x_2) = - \lambda_1\,{{\sigma_1}^3} -  \lambda_2\,{\cal K}^{(2)}  \, ,
\]
\bea 
Q_3(x_1,x_2,x_3) &=&  {\lambda_1}\,\left( {\sigma_1}\,{{{\sigma_2}}^4} + 
     {{{\sigma_1}}^4}\,{\sigma_2}\,{\sigma_3} + 
     {{\left(  4\,c^2 -1 \right) }
        \over {2\,\left( 1 + c \right) }} 
     \,{{{\sigma_1}}^2}\,
         {{{\sigma_2}}^2}\,{\sigma_3} - 
     {{3} \over {2\,\left( 1 + c \right) }} 
  \left(    \,{{{\sigma_2}}^3}\,{\sigma_3} + \,{{{\sigma_1}}^3}
       \,{{{\sigma_3}}^2} \right)
      \right)   \nonumber \\  & &
+ {\lambda_2}\,\left( {\sigma_1}\,{{{\sigma_2}}^4} + 
     {{{\sigma_1}}^4}\,{\sigma_2}\,{\sigma_3} - 
     2\,\left( 1 - c \right) \,
 \left( {{{\sigma_1}}^2}\,
         {{{\sigma_2}}^2}\,{\sigma_3}  - {\sigma_1}\,{\sigma_2}
       \,{{{\sigma_3}}^2}
 \right) - 
     {{{\sigma_2}}^3}\,{\sigma_3} - {{{\sigma_1}}^3}\,{{{\sigma_3}}^2}
     \right)  \nonumber \\   & &
+ {\lambda_3}\,  {\cal K}^{(3)}  \, ,\nonumber
\eea
where the residual kernel freedom of each solution
has been explicited\footnote{We do not 
report  here the general solution of $Q_4$ which already
contains an extremely large number of terms and is not particularly 
useful for the purposes of this work.}. 
Notice that in the above solutions  the trigonometrical dependence on
the coupling constant has been hidden in a simple rational dependence 
on the self--dual variable $c$ defined in  eq. (\ref{c}).
This major simplification has been made possible by noticing that
the systematic solution of the dynamical recursive equations alone
(\ref{dyn}) yields polynomials $Q_n$ which already have the correct
single--parameter  kernel ambiguity  (\ref{affine}) 
expected for the whole system. It therefore means that, 
actually the dynamical recursive equations (\ref{dyn}) are  
{\em equivalent} to the system 
of the two coupled equations (\ref{dyn})  and (\ref{kin}).

The general  solutions that we have found 
must include in particular
the form factors of the elementary field $\varphi (x)$ 
which were first studied in \cite{FMS2}.
One can prove  that they can in fact be selected  by imposing 
either the asymptotic vanishing of the form factors for large values of the 
rapidities (i.e. imposing the cancellation of the highest partial
degree terms in $Q_n$) or the proportionality $Q_n \sim \sigma_n$ \cite{FMS1}.
The $\lambda_i$ are then determined to be in this case 
\[ 
  \lambda_2^\varphi = -\lambda_1^\varphi \nonumber \:,
\]
\[
\lambda_i^\varphi = 0 \:\:\:\:\:\:\:\:\:\: \forall \,i>2\:.
\]
Finally the overall normalization is fixed by\footnote{Our convention
on the normalization of states is $\langle A(\th_1)| A(\th_2)\rangle =
2\pi \, \delta(\th_1-\th_2) = 2\pi E_1 \, \delta(p_1-p_2)$ .}
\[
\langle o | \varphi(0) | A \rangle = \frac{Z^{1/2}}{\sqrt{2}} \, ,
\]
which sets $\lambda_1^\varphi= \mu^{-1} Z^{1/2}/ \sqrt{2}$.
In the above expression $Z$ is the wave function renormalization 
constant of the theory which will be exactly computed in Section 6.

By using the above general solutions we can also identify
the 1--parameter family of the 
trace $\Theta(x)$ of the Stress--Energy tensor  for 
different values of the background charge. 
This family of operators  was studied in ref. \cite{MS} 
where the authors showed that 
different choices of $\Theta(x)$
select  different possible 
ultraviolet limits of the theory.
In order
to identify these form factors it is sufficient to impose the
proportionality $Q_n \sim \sigma_1\,\sigma_{n-1}$ for $n \geq 3$, as 
it can be shown from the  conservation of the Stress--Energy--Tensor.
In this way one determines all the free kernel parameters $\lambda_i$ but
the first two.  
The  parameter $\lambda_3^\Theta$ is  found to be for example 
\[
\lambda_3^\Theta = \lambda_2^\Theta + \frac{3 \, \lambda_1^\Theta}{2\,c+2} \:.
\] 
Finally, imposing the overall normalization 
\[
    F_2^\Theta (i\pi) = 2\, \pi\, m^2\, , 
\]
one determines 
\[
\lambda_2^\Theta = \frac{\pi \,m^2}{(c-1) \, \Gamma^2} \, ,
\]
and obtains a one--parameter family of independent operators 
for arbitrary $\lambda_1^\Theta$ which coincides with the one analyzed
in ref. \cite{MS}.

In order to identify different operators in this general space of
solutions one must resort to more powerful techniques.
We will see in the following section how the imposition 
of the cluster equations (\ref{cluster}) enables us to 
extract the form factors of a whole basis  
in the space of non--derivative scalar operators.

\resection{Form Factors of Exponential Operators}
In this chapter we  study the existence of solutions 
of the form factor equations 
which also satisfy the further requirement given by the  so--called 
cluster equations (\ref{cluster}) imposed on a multiparticle form
factor $F_n$.
This  restrictive set of non--linear equations 
is believed to select out the exponential operators 
in a Lagrangian theory \cite{Smi,KM}. More recently 
it has been shown in ref. \cite{DSC} 
that these equations are the distinguishing property 
of scaling operators in the conformal  limit of 
a two--dimensional field theory
at least in the cases where there is no symmetry preventing the 
form factors from being non--vanishing.
This observation has been confirmed and 
successfully employed in ref.'s
\cite{DS,AMV2} for identifying the complete set 
of scaling primary fields in some  
massive deformations of minimal models.
Cluster solutions become therefore objects of  
utmost interest in the BD model because
the two ways of looking at them
either as exponential operators or as scaling fields, 
converge in this theory where 
specific exponentials are identified
with primary operators in the reduced models
describing deformations of conformal field theories.

In order to impose the cluster equations (\ref{cluster})
we fix the overall normalization of the form factors 
by adopting the convenient choice $F_0 =1$ 
and choose
\[
Q_1 = 1 \:.
\]
Equations (\ref{cluster}) 
then amount to requiring the following property on the polynomial $Q_n$ 
\be \label{clQ}
\lim_{\Lambda\to \infty} \frac{ Q_n}{{\cal K}^{(n)}} (\Lambda\, x_1,\ldots,\Lambda\,
x_m,x_{m+1},\ldots,x_n) = 
t \,\, \frac{ Q_m}{{\cal K}^{(m)}} (\{x_i\}_{i=1,\ldots,m}) \:
\frac{ Q_{n-m}}{{\cal K}^{(n-m)}} (\{x_i\}_{i=m+1,\ldots,n}) \nonumber \:,
\ee
where $t$ -- the variable introduced in  eq. (\ref{H}) --
is now switched on and treated as
a free parameter.
These further restrictions imposed on the general solutions 
of residue equations determine level by level all the $\lambda_n$ parameters 
as functions of $t$. At any given level $n$, the number of equations 
which determine the only free parameter left $\lambda_n$, grows rapidly with $n$, 
therefore the very existence of a cluster solution is not at all 
obvious. For the first computed solutions however, all the equations
on a given $\lambda_n$ turn out to be identical and 
we believe that this should be the case at any level.
In this way we obtain a one--parameter family of solutions for $t$ arbitrary,
of which we report the first multiparticle
representatives\footnote{Notice that from a computational point of
view there is no difficulty in obtaining the next multiparticle
solutions since the dynamical recursive equations (\ref{dyn}) 
are linear equations in the unknown coefficients of independent monomials in
the $\sigma$'s and the dependence on the coupling constant is simply 
a rational dependence on $c$.}  
in Appendix A. 
Notice that 
$t$ is {\em not} an overall normalization factor
since the normalization of the form factors has been fixed 
by $F_0=1$ and indeed, 
due to the nonlinearity of (\ref{clQ}), 
the solutions $Q_n(t)$ turn out to be polynomials in $t$ of 
degree $n-1$.
This means that  $t$  defines through
the polynomials of Appendix A and 
eqs. (\ref{Fn}) and (\ref{H}) a one--parameter family 
of  solutions $F_n^{\{t\}}$ 
corresponding to independent operators.
If we make then the hypothesis that these solutions  actually 
correspond to the form factors of the exponential operators
$e^{kg\varphi (x)}$,
\be \label{norminvn}
F_n^{\{t\}}(\th_1,\ldots,\th_n) = 
\frac{\langle 0| e^{kg\varphi(0)} |A(\th_1) \cdots A(\th_n) \rangle}
                   {\langle 0| e^{kg\varphi(0)} |0 \rangle} \, ,
\ee
we are  forced to consider  $t$ 
 as  a well--defined function $t(k,B)$ of 
$k$ and $B$ rather than a free parameter.
In particular, in order to establish the one--to--one correspondence
between cluster solutions and exponential operators it
is of particular interest to compute the normalization--invariant
quantity 
\be \label{norminv}
F_1^{\{t\}} = \frac{\langle 0| e^{kg\varphi(0)} |A \rangle}
                   {\langle 0| e^{kg\varphi(0)} |0 \rangle} 
            =  \mu(B) t(k,B) \:.
\ee
In the following we  consider in detail some conditions that 
we can impose on the function $t(k,B)$ in order to find its exact 
form.

\subsection{The Function $t(k,B)$}
The first information on $t(k,B)$ can be obtained from 
the computation of the conformal dimensions \linebreak
$\Delta = - g^2\, k^2 /8\,\pi$ of the 
operators $e^{kg\varphi (x)}$   in the free--boson 
ultraviolet limit at lowest order in $g^2$. 
These can be easily obtained from the analysis of the 
short distance behavior of the correlator $\langle 0| e^{kg\varphi (x)} e^{kg\varphi (0)}|0 \rangle$
by means of eq. (\ref{spectral}) and the cluster solutions $F_n^{\{t\}}$ .
We obtain
\[
\Delta = - g^2\,\lim_{g\to 0}\frac{ \mu(B)^2\, t(k,B)^2}{4\, \pi\, g^2} = 
          - \frac{g^2\, t(k,0)^2}{8 \,\pi} \:,
\]
from which one obtains the important relation
\be \label{tk0}
\lim_{B\to 0} t(k,B)= k \:.
\ee
Furthermore, from the expressions (\ref{Q3}) and (\ref{Q4}), by imposing the 
proportionality $Q_n \sim \sigma_1\,\sigma_{n-1}$,
 one can easily verify that the only cluster solutions
which also belong to the class of possible traces of the
stress--energy tensor are defined by the solutions of
\[ 
-1 + 2\,c + 2\,t + 2\,c\,t + 2\,{t^2} + 2\,c\,{t^2} = 0 \:,
\]
namely 
\be \label{tpm}
t^\pm= 
 \left\{ 
 \begin{array}{l} 
\bd \frac{ \sin ((B+1)\pi/6)}{\cos ((B+2)\pi/6)} \ed\\
\nonumber \\ 
\bd \frac{ \sin ((B-3)\pi/6)}{\cos ((B+2)\pi/6)}  \ed     
    \end{array} \right.\:.
\ee
These two solutions correspond 
to the ones found in ref. \cite{MS}
and  identified with the form factors of the fundamental vertex operators
$e^{g\varphi}$ and $e^{-2g\varphi}$ which appear in the Lagrangian density.
This can also be obtained immediately by 
taking the limit $B\to 0$ in eq. (\ref{tpm}) 
which gives respectively $k=1,-2$ in virtue of (\ref{tk0}).
Therefore we have also the two  following important requirements on $t(k,B)$:
\be
 \label{t1B} t(1,B) = \frac{ \sin ((B+1)\pi/6)}{\cos ((B+2)\pi/6)}  \,,
\ee 
\be
 \label{t-2B} t(-2,B) =  \frac{ \sin ((B-3)\pi/6)}{\cos ((B+2)\pi/6)}  \,.
\ee
As a limiting case of the cluster solutions we can also recover  
the form factors of
the fundamental field $\varphi(x)$ which is naturally obtained from the vertex
operators in the limit $k\to 0$. These form factors of course 
satisfy a trivial  
cluster property because they vanish for large rapidities 
and therefore satisfy  eq. (\ref{clQ})  with $t=0$.
Hence we get one more information
\be \label{t0B}
 \lim_{k\to 0} t(k,B) = 0 \:.
\ee
Indeed one can easily check that the form factors 
we had obtained in the previous section 
for the field $\varphi(x)$ from the most general solutions
of residue equations satisfy, 
\[
  F_n^{\varphi} = \lambda_1^\varphi \: \lim_{t\to 0} \frac{F_n^{\{t\}}}{t} \:.
\]
A   remarkable check on the correct identification of these operators 
is obtained studying the quantum equations of motion of the model
\[
 \square \varphi  + \frac{m_0^2}{3\,g} \,
\left( \, e^{g\varphi} -  e^{-2g\varphi} \right)= 0 \, .
\]
If our identification is correct we should find\footnote{In general
$F_n^{\square\Phi} = - m^2 \frac{\sigma_1 \sigma_{n-1}}{\sigma_n}
F_n^{\Phi}$ for any field $\Phi(x)$.}
\[
m^2\,\frac{\sigma_1^{(n)}\,\sigma_{n-1}^{(n)} }{\sigma_n^{(n)}} \, F_n^\varphi + \tau \, 
\left(  F_n^{\{t^+\}} - F_n^{\{t^-\}}\right) = 0 \, , 
\]
with some constant $\tau$, or equivalently  
\[
\lambda_1^\varphi \,m^2\,\frac{\sigma_1^{(n)}\,\sigma_{n-1}^{(n)} }{\sigma_n^{(n)}} \, Q_n(0) + \tau \, 
\left( t^+ \, Q_n(t^+) - t^- \, Q_n(t^-) \right) = 0 \, .
\]
Indeed this last equation can be  verified to hold 
on the solutions given in Appendix A with 
\[
\tau = - \frac{\lambda_1^\varphi m^2}{\sqrt{3}} \, \tan ((B+2) \pi/6).
\]
The   non--perturbative nature of this last check  
shows that the identification of cluster 
solutions as vertex operators
is far beyond a semiclassical one for small coupling constant.

The constraints obtained for the function $t(k,B)$, eqs. (\ref{tk0}), 
(\ref{t1B}), (\ref{t-2B}) and (\ref{t0B}) are 
not sufficient to determine its 
form and, in particular, 
little information is given on the  dependence on $k$.
We will see however in the next section that some additional 
requirements coming from the
reductions of the ZMS model impose a periodicity condition 
in  $k$ for the function $t(k,B)$
\be \label{periodic}
t(k,B)= t(k+6/B,B)\, ,
\ee
which suggests the following conjecture:
\be \label{ansatz}
t(k,B) = \frac{\sin (k\,B\,\pi/6) \,\, \sin ((k\,B+ B + 2)\pi/6) }
              {2\,\sin (B\,\pi/6) \,\,\sin ((2-B)\pi/6) \,\, 
              \cos ((B+2)\pi/6) } \, .
\ee
This function  satisfy all the aforementioned requirements.
A decisive check of the validity of this expression  
will be  obtained in the following chapter by 
the comparison with explicit computations 
of  form factors of primary operators 
in specific reductions of the ZMS model.
This formula may be regarded as one of the main results of the paper.
In fact it allows us  to explicitly assign to every vertex operator
$e^{kg\varphi}$ in the BD model its  form factors $F_n^{\{k\}}$ 
which are obtained from the cluster solutions  $Q_n(t)$ of Appendix A 
through the parameterization  (\ref{Fn}) and 
eq. (\ref{H})  by replacing $t=t(k,B)$.

\resection{Form Factors in the reductions of the ZMS model}
We now turn our attention to the analitical continuation of the model
to imaginary values of the coupling constant $g$, 
namely to possible reductions of the ZMS model.
In these models the spectrum  is no more a single--particle 
one as in the real coupling BD model, but it has a richer structure that
depends on the model analyzed. 
We consider here only those restrictions whose spectrum 
still contains the elementary boson excitation 
of the BD model, namely $\phi_{1,2}$ and some cases of $\phi_{1,5}$
deformations\footnote{To avoid confusion we stress that the elementary
BD scalar boson, which is created from the field $\varphi(x)$, is not 
the fundamental particle in the bootstrap of the 
reductions of the ZMS model which is instead a three--component kink.}.
If we assume  that the identification obtained between cluster solutions 
and vertex operators of the model is exact, we are then 
led to establish a correspondence between 
the form factors of exponential operators $e^{kg\varphi(x)}$ 
in the BD model and the form 
factors\footnote{The form factors in the reduced model 
must be intended as the matrix elements on the BD breather sector.}
of scaling primary operators in the 
deformations according to the correspondence given by eq. 
(\ref{map}) and  Table 1.
An immediate  consistency  requirement for  
this procedure is obtained by imposing 
that the form factors respect the symmetry (\ref{symkac}) 
of the Kac table of minimal models. 
For example, the quantity $F_1^{\{t\}}$ of eq. (\ref{norminv}) 
should have the same value if evaluated at 
$k=k_{m,n}$ and $k=k_{r-m,s-n}$.
Imposing this condition both in the $\phi_{1,2}$ deformations and in
the $\phi_{1,5}$ relevant ones
we obtain respectively that the following two 
symmetries of the function $t(k,B)$ must hold
\be\ba{rcl}
t(k,B) &=& t(-k-1-2/B,B)\, , \\
t(k,B) &=& t(-k-1+4/B,B)\, , 
\ea\ee
which in particular entail the  above mentioned periodicity in $k$, equation
(\ref{periodic}).
Both these symmetries are indeed separately  
satisfied by the function (\ref{ansatz}).

A  precise check on the validity of equation 
(\ref{ansatz}) is provided  by comparing its predictions 
with the form factors of scaling primary operators in $\phi_{1,2}$ and 
$\phi_{1,5}$ deformations which can be found in literature.
We have indeed computed the normalization invariant 
ratio $F_1/F_0$  using eq. (\ref{norminv})
and the assignments of Table 1,
for all the known cases of primary form factors 
which have been analyzed in literature
\cite{YL,DM,DS,AMV,AMV2} (see Table 2)  
and a perfect agreement has been found with all the values
reported in the references.
We stress here the fact that in the references considered, 
the form factors of primary operators have been identified by different 
techniques: in ref.'s \cite{YL,DM,AMV} the identification 
has been obtained by using the correspondence 
between the deforming field and the trace of the stress--energy tensor
whereas in ref.'s \cite{DS,AMV2} the form factors of the primary fields
have been identified with  the finite number of solutions
of a non--linear system of cluster equations 
involving the form factors relative to  
the whole particle spectrum of the reduced models. 

\resection{The Wave Function Renormalization Constant of the BD Model
and the Form Factors of $\varphi(x)$ and $:\!\varphi^2(x)\!:$}
The  form factors $F_n^{\{k\}}(\th_1,\ldots,\th_n)$ that we have
computed  have been so far conveniently normalized putting
$F_0=1$. From equation (\ref{norminvn}) one immediately observes
that these form factors are invariant under an additive redefinition
of the field $\varphi(x)\to \varphi(x) + const$. We  remove this
ambiguity on the definition of the field $\varphi(x)$ by imposing that
its vacuum expectation value $\langle 0| \varphi(x) |0
\rangle$ be zero, namely subtracting from the original Lagrangian field 
the value of the one point tadpole function.
Consider now the following expansion of the form factors of 
exponential operators:
\[
\langle 0| e^{kg\varphi(0)} |A(\th_1) \cdots A(\th_n) \rangle = 
\sum_{j=1}^{\infty} \, \frac{k^j\,g^j}{j!}
\langle 0| :\!\varphi^j(0)\!: |A(\th_1) \cdots A(\th_n) \rangle \, ,
\]
and of the vacuum expectation value
\[
\langle 0| e^{kg\varphi(0)} |0 \rangle = 
\sum_{j=0}^{\infty} \, \frac{k^j\,g^j}{j!}
\langle 0| :\!\varphi^j(0)\!: |0 \rangle = 1 + o(k^2) \, .
\]
If we now expand the form factors $F_n^{\{k\}}$ that we have 
obtained\footnote{Here and in the following we will adopt
the notation $F_n^{\{k\}}$ instead of $F_n^{\{t\}}$ to stress 
the dependence on $k$. The relation between the two expressions is 
obviously given by $t=t(k,B)$ eq. (\ref{ansatz}).} in series of
$k$ we can identify the form factors of the fields $\varphi(x)$ and
$:\!\varphi^2(x)\!:$ as the coefficients of order $k$ and $k^2$ 
respectively. 
\be \label{expansion}
\ba{rcl}
F_n^{\{k\}}(\th_1,\ldots,\th_n) &=& \bd
\frac{\langle 0| e^{kg\varphi(0)} |A(\th_1) \cdots A(\th_n) \rangle}
                   {\langle 0| e^{kg\varphi(0)} |0 \rangle} \ed\\
&=& \nonumber \bd
 k\, g\, \langle 0| \varphi(x) |A(\th_1) \cdots A(\th_n) \rangle  +
\frac{k^2 \, g^2}{2}\, \langle 0| :\!\varphi^2(x)\!: |A(\th_1) 
\cdots A(\th_n) \rangle  + o(k^3) \, .\ed
\ea
\ee
This procedure gives the form factors of  
$\varphi(x)$ and $:\!\varphi^2(x)\!:$ with the correct overall
normalization of the fields. This observation in
particular allows the exact determination of the wave function
renormalization constant $Z(B)$ of the BD model. In fact, considering the 
first order expansion in $k$ of $F_1^{\{k\}}$ 
\[
\ba{rcl} 
F_1^{\{k\}} &=& \mu(B) \, t(k,B) = \mu(B)\bd \frac{k\,B\, \pi
\tan((B+2)\pi/6)}{12 \,\sin (B \pi/6)\,\sin ((2-B) \pi/6) } \ed+ o(k^2) 
\nonumber \\
&=& k\, g \, \langle 0 |\varphi(0) |A\rangle + o(k^2) \nonumber
\\
&=& \bd \frac{k\,g\,Z^{1/2}}{\sqrt{2}}\,  + o(k^2) \ed \, ,\nonumber
\ea
\]
one easily obtains the following expression for $Z(B)$ 
\bea
Z(B) &=& \mu(B)^2 \,B\,(2-B) \, \frac{\pi}{288} 
\left(
\frac{\tan((B+2)\pi/6)}{\sin (B \pi/6)\,\sin ((2-B) \pi/6)}
\right)^2 \\
&=& \frac{2\, \pi}{3\, \sqrt{3}}  \,\frac{ B\,(2-B)}{{\cal N} (B)} \,
\frac{(c-1)}{(1+2\,c)\,(1-2\,c)} \, , \nonumber
\eea
where ${\cal N}(B)$ is defined in eq. (\ref{N}).
The function $Z(B)$ is manifestly dual with respect to the
weak--strong coupling transformation $B\longleftrightarrow 2-B$ and 
can be easily shown to coincide  at lowest order in $g^2$ with 
the correct perturbative result coming from the 
one--loop self energy diagram
\[
Z = 1 - \frac{g^2}{12} \, \left( \frac{1}{\pi} - \frac{1}{3\sqrt{3}}
\right) + o(g^4) \,.
\]
A plot of the function $Z(B)$ is given in Figure 2.
Notice the tiny deviation of the constant from the free field
value $Z=1$ on the entire range of the coupling constant $B \in [0,2]$.

The correctly normalized form factors of the field $\varphi(x)$ 
are given by 
\bea
{F_n}^\varphi &=& g^{-1} \, \left. \frac{d}{d\,k} F_n^{\{k\}} \right|_{k=0} \\
            &=& \nonumber \frac{Z^{1/2}}{\mu \sqrt{2}} \,
   \left. \frac{F_n^{\{t\}}}{t} \right|_{t=0} \, ,
\eea
while the exact form factors of the field $:\!\varphi^2 (x)\!:$ are simply
obtained by 
\be
{F_n}^{\varphi^2} = g^{-2} \, 
\left. \frac{d^2}{d\,k^2} F_n^{\{k\}} \right|_{k=0} \, ,
\ee
For example we can compute 
\[
\ba{rcl}
{F_1}^{\varphi^2} &=& \langle 0 | :\!\varphi^2 (0)\!: | A \rangle  \\ 
 &=& \bd  \mu(B) \, g^{-2} \, \left. \frac{d^2}{dk^2} \,t(k,B)
 \right|_{k=0}  \ed \\
&=& \bd \mu(B) \, B\, (2-B) \, \frac{\pi}{144} 
\,\frac{1}{\sin (B \pi/6)\,\sin ((2-B) \pi/6)}   \, ,               
\ed \ea
\]
which exactly matches at lowest order in $g$ 
with the one loop calculation
\[
\langle 0 | :\!\varphi^2 (0)\!: | A \rangle = \frac{g}{6\, \sqrt{6}}  + o(g^3).
\]
In a similar way we get 
\[\ba{rcl}
{F_2}^{\varphi^2} (\th_1-\th_2) &=& \bd
\langle 0 | :\!\varphi^2 (0)\!: | A(\th_1)\,A(\th_2) \rangle \ed\\
&=& \bd \nonumber \mu^2(B) \, B \, (2-B) \, \frac{\pi}{288} \,
\frac{1}{(\sin (B \pi/6)\,\sin ((2-B) \pi/6))^2} \, \ed\\ 
& & \bd \nonumber \cdot\Big( \sigma_1^3 \, \tan^2 ((B+2)\pi/6)  - 
\sigma_1 \sigma_2 \big( 2\,\sin (B \pi/6)\,\sin ((2-B) \pi/6) 
 + \tan^2 ((B+2)\pi/6)\big) \Big) \ed \\
& & \bd 
\cdot \frac{F^{min}(\th_{1}-\th_{2})}
{(x_2 + x_2)(x_1^2  + x_1 x_2 +x_2^2)} \nonumber \,. \ed
\ea
\]
Notice that in  order to obtain the form factors of arbitrary 
operators $:\!\varphi^n(x)\!:$ one should exactly 
compute the vacuum expectation value
$\langle 0 | e^{kg\varphi(0)}|0 \rangle$  
of the exponential operators and make use of expansion 
(\ref{expansion}) 
(for the sine--Gordon model the vacuum expectation value
of the exponential operators has been recently obtained
in ref. \cite{LZ}).

\resection{Conclusions}
In this paper we have computed, in the framework of 
the bootstrap approach to integrable models,
the first multiparticle solutions of form factor equations 
for general non--derivative scalar operators in the BD model. 
Among these solutions we have selected a one--parameter
family of cluster solutions which have been identified
by means of the central result eq. (\ref{ansatz}) 
with the form factors of exponential operators $e^{kg\varphi}$.
In the complex coupling constant version of the model,
the form factors of exponential operators 
allow to identify the  form factors of relevant 
primary operators in the  sector of the lightest breather of 
$\phi_{1,2}$ and $\phi_{1,5}$ deformations of minimal models
and perfect agreement has been found with all the examples that we
have found in literature.
Finally, by using  the cluster solutions, we have 
computed the form factors of the fields $\varphi(x)$ and 
$:\!\varphi^2(x)\!:$ with the correct overall normalization 
and determined in this way the non--perturbative exact
wave function renormalization constant of the model.

We have therefore  obtained 
the characterization of form factors for a whole basis
in the space of scalar 
non--derivative operators 
and we have found complete consistency,
in a  non--perturbative setting,
between the axiomatic $S$--matrix approach
to bootstrap systems and the Lagrangian 
approach to quantum field theories.

This work also yields an efficient tool 
for the identification of relevant primary 
fields among the cluster solutions of 
massive $\phi_{1,2}$ and $\phi_{1,5}$ deformations
of minimal models.
\newpage

\noindent{\large \bf Acknowledgments:}
I wish to express my grateful thanks to  G. Mussardo
for his continuous and warm support in the preparation of this paper.
I want to thank also G. Delfino, A. De Martino and F. Smirnov for
precious discussions.
I am grateful to the INFN for supporting
a monthly stay in Paris at Institute Henry Poincar\'e where the work
has been completed and to prof. O. Babelon for his ospitality.

\newpage 
 
\appendix
\appsection

In this Appendix we list the first solutions of the one--parameter 
family of $Q_n$ polynomials of 
cluster solutions in the BD model. 
In the following expressions, the variable $c$ 
is the dual--invariant function of the coupling constant
defined in eq. (\ref{c}) and $t$ is a free parameter.
The solutions are identified with those of the basis of 
operators $e^{kg\varphi}$ by means of eq. (\ref{ansatz}) 
which determines  $t$ as a function of $k$ and $g$.
\be
Q_1(t) = 1\,,
\ee
\begin{eqnarray}
Q_2(t) &=& t\,{\sigma_1^3}  \\
   \nn & & - \left( 1 + t \right) \,\sigma_1\, \sigma_2 \,,
\end{eqnarray}
\begin{eqnarray} \label{Q3}  
Q_3(t) \,\, \,2 \,(1+c)  &=&   
2\,\left( 1 + c \right) \,{t^2}\,{{\sigma_1}^3}\,{{\sigma_2}^3} \\ 
\nn & & - 2\,\left( 1 + c \right) \,t\,\left( 1 + t \right) \,\sigma_1\,
   {{\sigma_2}^4} \\ 
\nn & &- 2\,\left( 1 + c \right) \,t\,\left( 1 + t \right) \,
   {{\sigma_1}^4}\,\sigma_2\,\sigma_3 \\ 
\nn & &+ 
  \left( 3 + 4\,t - 4\,{c^2}\,t - 2\,{t^2} - 2\,c\,{t^2} \right) \,
   {{\sigma_1}^2}\,{{\sigma_2}^2}\,\sigma_3 \\ 
\nn & &+ 
  \left( -1 + 2\,c + 2\,t + 2\,c\,t + 2\,{t^2} + 2\,c\,{t^2} \right) \,
   {{\sigma_2}^3}\,\sigma_3 \\ 
\nn & &+ 
  \left( -1 + 2\,c + 2\,t + 2\,c\,t + 2\,{t^2} + 2\,c\,{t^2} \right) \,
   {{\sigma_1}^3}\,{{\sigma_3}^2} \\ 
\nn & &+ 
  4\,\left( -1 + c \right) \,\left( 1 + c \right) \,\left( 1 + t \right) \,
   \sigma_1\,\sigma_2\,{{\sigma_3}^2}\,,
\end{eqnarray}
\begin{eqnarray}\label{Q4}
 & & \!\!\!\!\!\!\!\! Q_4(t)\,\, \,2\, (1+c) =\\ \nn &=& 
 2\,\left( 1 + c \right) \,{t^3}\,{{\sigma_1}^3}\,{{\sigma_2}^3}\,
   {{\sigma_3}^3}\\ 
\nn & & - 2\,\left( 1 + c \right) \,{t^2}\,
   \left( 1 + t \right) \,\sigma_1\,{{\sigma_2}^4}\,
   {{\sigma_3}^3}\\ 
\nn & & - 2\,\left( 1 + c \right) \,{t^2}\,
   \left( 1 + t \right) \,{{\sigma_1}^4}\,\sigma_2\,
   {{\sigma_3}^4} \\ 
\nn & &+ t\,\left( 3 + 4\,t - 4\,{c^2}\,t - 2\,{t^2} - 
     2\,c\,{t^2} \right) \,{{\sigma_1}^2}\,{{\sigma_2}^2}\,
   {{\sigma_3}^4}\\ 
\nn & & + t\,\left( -1 + 2\,c + 2\,t + 2\,c\,t + 2\,{t^2} + 
     2\,c\,{t^2} \right) \,{{\sigma_2}^3}\,{{\sigma_3}^4} \\ 
\nn & &+ 
  t\,\left( -1 + 2\,c + 2\,t + 2\,c\,t + 2\,{t^2} + 2\,c\,{t^2} \right) \,
   {{\sigma_1}^3}\,{{\sigma_3}^5} \\ 
\nn & &+ 
  4\,\left( -1 + c \right) \,\left( 1 + c \right) \,t\,\left( 1 + t \right) \,
   \sigma_1\,\sigma_2\,{{\sigma_3}^5} \\ 
\nn & &- 
  2\,\left( 1 + c \right) \,{t^2}\,\left( 1 + t \right) \,
   {{\sigma_1}^3}\,{{\sigma_2}^4}\,\sigma_3\,\sigma_4 \\ 
\nn & &
   + 2\,\left( 1 + c \right) \,t\,{{\left( 1 + t \right) }^2}\,
   \sigma_1\,{{\sigma_2}^5}\,\sigma_3\,\sigma_4 \\ 
\nn & &+ 
  t\,\left( 3 + 4\,t - 4\,{c^2}\,t - 2\,{t^2} - 2\,c\,{t^2} \right) \,
   {{\sigma_1}^4}\,{{\sigma_2}^2}\,{{\sigma_3}^2}\,
   \sigma_4 \\ 
\nn & &+ 2\,\left( 1 + t \right) \,
   \left( -2 + c - 2\,t + 2\,c\,t + 4\,{c^2}\,t + 3\,{t^2} + 3\,c\,{t^2}
      \right) \,{{\sigma_1}^2}\,{{\sigma_2}^3}\,
   {{\sigma_3}^2}\,\sigma_4 \\ 
\nn & &+ 
  \left( 1 + t \right) \,\left( 1 - 2\,c - 2\,t - 2\,c\,t - 2\,{t^2} - 
     2\,c\,{t^2} \right) \,{{\sigma_2}^4}\,{{\sigma_3}^2}\,
   \sigma_4 \\ 
\nn & &+ t\,\left( -1 + 2\,c + 2\,t + 2\,c\,t + 2\,{t^2} + 
     2\,c\,{t^2} \right) \,{{\sigma_1}^5}\,{{\sigma_3}^3}\,
   \sigma_4 \\ 
\nn & &+ \left( 1 - 2\,c - 4\,t + 4\,{c^2}\,t - 2\,{t^2} + 
     14\,c\,{t^2} + 8\,{c^2}\,{t^2} - 8\,{c^3}\,{t^2} + 6\,{t^3} + 6\,c\,{t^3}
      \right) \,{{\sigma_1}^3}\,\sigma_2\,{{\sigma_3}^3}\,
   \sigma_4 \\ 
\nn & &+ \left( 7 - 8\,c + 9\,t - 14\,c\,t - 12\,{c^2}\,t + 
     8\,{c^3}\,t - 2\,{t^2} - 6\,c\,{t^2} - 4\,{c^2}\,{t^2} - 2\,{t^3} - 
     2\,c\,{t^3} \right) \,\sigma_1\,{{\sigma_2}^2}\,
   {{\sigma_3}^3}\,\sigma_4 \\ 
\nn & &+ 
  t\,\left( 3 - 14\,c + 8\,{c^3} - 6\,t - 14\,c\,t + 8\,{c^3}\,t - 6\,{t^2} - 
     6\,c\,{t^2} \right) \,{{\sigma_1}^2}\,{{\sigma_3}^4}\,
   \sigma_4 \\ 
\nn & &+ 2\,\left( -1 + c \right) \,
   \left( 1 - 2\,c - 2\,t - 2\,c\,t - 2\,{t^2} - 2\,c\,{t^2} \right) \,
   \sigma_2\,{{\sigma_3}^4}\,\sigma_4 \\ 
\nn & &+ 
  t\,\left( -1 + 2\,c + 2\,t + 2\,c\,t + 2\,{t^2} + 2\,c\,{t^2} \right) \,
   {{\sigma_1}^4}\,{{\sigma_2}^3}\,{{\sigma_4}^2} \\ 
\nn & &+ 
  \left( 1 + t \right) \,\left( 1 - 2\,c - 2\,t - 2\,c\,t - 2\,{t^2} - 
     2\,c\,{t^2} \right) \,{{\sigma_1}^2}\,{{\sigma_2}^4}\,
   {{\sigma_4}^2} \\ 
\nn & &+ 4\,\left( -1 + c \right) \,\left( 1 + c \right) \,
   t\,\left( 1 + t \right) \,{{\sigma_1}^5}\,\sigma_2\,
   \sigma_3\,{{\sigma_4}^2} \\ 
\nn & &+ 
  \left( 7 - 8\,c + 9\,t - 14\,c\,t - 12\,{c^2}\,t + 8\,{c^3}\,t - 2\,{t^2} - 
     6\,c\,{t^2} - 4\,{c^2}\,{t^2} - 2\,{t^3} - 2\,c\,{t^3} \right) \,
   {{\sigma_1}^3}\,{{\sigma_2}^2}\,\sigma_3\,
   {{\sigma_4}^2} \\ 
\nn & &+ \left( -5 + 14\,c - 8\,{c^2} - 6\,t - 2\,c\,t + 
     4\,{c^2}\,t - 6\,{t^2} - 14\,c\,{t^2} + 8\,{c^3}\,{t^2} - 4\,{t^3} - 
     4\,c\,{t^3} \right) \,\sigma_1\,{{\sigma_2}^3}\,
   \sigma_3\,{{\sigma_4}^2} \\ 
\nn & &+ 
  t\,\left( 3 - 14\,c + 8\,{c^3} - 6\,t - 14\,c\,t + 8\,{c^3}\,t - 6\,{t^2} - 
     6\,c\,{t^2} \right) \,{{\sigma_1}^4}\,{{\sigma_3}^2}\,
   {{\sigma_4}^2} \\ 
\nn & &+ 2\,\left( -5 + 14\,c - 8\,{c^2} - 2\,t + 8\,c\,t - 
     6\,{c^2}\,t - 8\,{c^3}\,t + 8\,{c^4}\,t + {t^2} - 11\,c\,{t^2} - 
     4\,{c^2}\,{t^2} \right.\\
   \nn & & \:\:\:\:\:\:\:\:\:\:\: \left.
     + 8\,{c^3}\,{t^2} - 3\,{t^3} - 3\,c\,{t^3} \right) \,
   {{\sigma_1}^2}\,\sigma_2\,{{\sigma_3}^2}\,
   {{\sigma_4}^2} \\ 
\nn & & + \left( 1 + 4\,c - 4\,{c^2} + 2\,t \right) \,
   \left( -1 + 2\,c + 2\,t + 2\,c\,t + 2\,{t^2} + 2\,c\,{t^2} \right) \,
   {{\sigma_2}^2}\,{{\sigma_3}^2}\,{{\sigma_4}^2} \\ 
\nn & & + 
  \left( -4\,c + 12\,{c^2} - 8\,{c^3} - 3\,t + 22\,c\,t - 16\,{c^3}\,t + 
     6\,{t^2} + 22\,c\,{t^2} - 16\,{c^3}\,{t^2} + 6\,{t^3} + 6\,c\,{t^3}
      \right) \,\sigma_1\,{{\sigma_3}^3}\,{{\sigma_4}^2} \\ 
\nn & & + 
  2\,\left( -1 + c \right) \,\left( 1 - 2\,c - 2\,t - 2\,c\,t - 2\,{t^2} - 
     2\,c\,{t^2} \right) \,{{\sigma_1}^4}\,\sigma_2\,
   {{\sigma_4}^3} \\ 
\nn & & + \left( 1 + 4\,c - 4\,{c^2} + 2\,t \right) \,
   \left( -1 + 2\,c + 2\,t + 2\,c\,t + 2\,{t^2} + 2\,c\,{t^2} \right) \,
   {{\sigma_1}^2}\,{{\sigma_2}^2}\,{{\sigma_4}^3}\\ 
\nn & &  + 
  \left( -4\,c + 12\,{c^2} - 8\,{c^3} - 3\,t + 22\,c\,t - 16\,{c^3}\,t + 
     6\,{t^2} + 22\,c\,{t^2} - 16\,{c^3}\,{t^2} + 6\,{t^3} + 6\,c\,{t^3}
      \right) \,{{\sigma_1}^3}\,\sigma_3\,{{\sigma_4}^3}\\ 
\nn & &  + 
  \left( 9 - 30\,c + 20\,{c^2} + 16\,{c^3} - 16\,{c^4} + 8\,t - 16\,c\,t + 
     8\,{c^2}\,t + 16\,{c^3}\,t - 16\,{c^4}\,t  \right.\\
    \nn & & \:\:\:\:\:\:\:\:\:\:\: \left.   + 
     2\,{t^2} + 10\,c\,{t^2} - 
     8\,{c^3}\,{t^2} + 2\,{t^3} + 2\,c\,{t^3} \right) \,\sigma_1\,
   \sigma_2\,\sigma_3\,{{\sigma_4}^3}\\ 
\nn & &  + 
  \left( 4\,c - 4\,{c^2} + t \right) \,
   \left( 1 - 2\,c - 2\,t - 2\,c\,t - 2\,{t^2} - 2\,c\,{t^2} \right) \,
   {{\sigma_3}^2}\,{{\sigma_4}^3} \\ 
\nn & & + 
  \left( 4\,c - 4\,{c^2} + t \right) \,
   \left( 1 - 2\,c - 2\,t - 2\,c\,t - 2\,{t^2} - 2\,c\,{t^2} \right) \,
   {{\sigma_1}^2}\,{{\sigma_4}^4} \,.
\end{eqnarray}

\newpage 
\vspace{25mm}

\noindent {\bf Table Captions}

\vspace{1cm}

\begin{description} 
\item [Table 1] Complex Liouville Theory assignments between
                 exponential operators and primary fields
                 for different choices of the screening operator.
\item [Table 2] Primary operators in ZMS reduced models 
  for which the form factors have been computed in literature.
\end{description}

\begin{center}
\vspace{1cm}
\begin{tabular}{|c|c|c|c|} \hline
\rule[-2mm]{0mm}{7mm} {\it Screening operator}& \multicolumn{1}{c|}{\it Deformation}
                                  & \multicolumn{1}{c|}{\it $B$}  
                                  & \multicolumn{1}{c|}{\it $k_{m,n}$}\\ \hline 
\rule[-2mm]{0mm}{7mm}$e^{-2g\varphi}$    & 
$ e^{g\varphi} = \phi_{1,2}$ & $\frac{2\,r}{r-2\,s}$  & $(n-1)-(m-1)\frac{s}{r}$\\  
\rule[-2mm]{0mm}{7mm}$e^{-2g\varphi}$    &
 $ e^{g\varphi} = \phi_{2,1}$ & $\frac{2\,s}{s-2\,r}$  & $(m-1)-(n-1)\frac{r}{s}$\\  
\rule[-2mm]{0mm}{7mm}$e^{g\varphi}$      &
$ e^{-2g\varphi} = \phi_{1,5}$ & $\frac{4\,r}{2\,r-s}$  & $\frac{1}{2}\left(
(1-n)-(1-m)\frac{s}{r}\right) $\\  
\rule[-2mm]{0mm}{7mm}$e^{g\varphi}$    & 
$ e^{-2g\varphi} = \phi_{5,1}$ & $\frac{4\,s}{2\,s-r}$  & $\frac{1}{2}\left(
(1-m)-(1-n)\frac{r}{s}\right) $\\  \hline
\end{tabular}
\vspace{.2cm}
\end{center}
\begin{center}
{\bf Table 1}
\end{center}
\begin{center}
\vspace{1cm}
\begin{tabular}{|c|c|c|c|c|} \hline
\rule[-2mm]{0mm}{7mm} {\it Model}& \multicolumn{1}{c|}{\it Deformation} & \multicolumn{1}{c|}{\it Primaries analyzed} &
 \multicolumn{1}{c|}{\it $F_1/F_0$} & \multicolumn{1}{c|}{\it Reference}  \\ \hline 
\rule[-2mm]{0mm}{7mm}$ {\cal M}_{2,5}$ &$\phi_{1,2}$   & $\phi_{1,2}$ & $0.8372182 \,i $ & \cite{YL} \\ 
\rule[-2mm]{0mm}{7mm}$ {\cal M}_{2,7}$ &$\phi_{1,2}$   & $\phi_{1,2}$ & $0.8129447\, i $ & \cite{AMV2}\\ 
\rule[-2mm]{0mm}{7mm}                  &               & $\phi_{1,3}$ & $1.245504\, i $ & \cite{AMV2}\\ 
\rule[-2mm]{0mm}{7mm}$ {\cal M}_{2,9}$ &$\phi_{1,2}$   & $\phi_{1,2}$ & $0.7548302 \, i $ & \cite{AMV2}\\ 
\rule[-2mm]{0mm}{7mm}                  &               & $\phi_{1,3}$ & $1.288576 \, i $ & \cite{AMV2}\\
\rule[-2mm]{0mm}{7mm}                  &               & $\phi_{1,4}$ & $1.564863\, i  $ & \cite{AMV2}\\
\rule[-2mm]{0mm}{7mm}$ {\cal M}_{3,4}$ &$\phi_{1,2}$   & $\phi_{1,2}$ & $-0.6409021 $ & \cite{DM,DS}\\ 
\rule[-2mm]{0mm}{7mm}                  &               & $\phi_{2,1}$ & $-3.706584 $ &\cite{DS} \\ 
\rule[-2mm]{0mm}{7mm}$ {\cal M}_{4,5}$ &$\phi_{1,2}$   & $\phi_{1,2}$ & $-0.8113145 $ & \cite{AMV}\\ 
\rule[-2mm]{0mm}{7mm}$ {\cal M}_{6,7}$ &$\phi_{1,2}$   & $\phi_{1,2}$ & $-0.9499626 $ & \cite{AMV}\\ 
\rule[-2mm]{0mm}{7mm}$ {\cal M}_{2,9}$ &$\phi_{1,4}\equiv\phi_{1,5}$    & $\phi_{1,2}$ & $-0.5483649 $ & \cite{AMV2}\\ 
\rule[-2mm]{0mm}{7mm}                  &               & $\phi_{1,3}$ & $-1.476188 $ & \cite{AMV2}\\ 
\rule[-2mm]{0mm}{7mm}                  &               & $\phi_{1,4}$ & $-2.169493 $ & \cite{AMV2}\\ \hline
\end{tabular}
\vspace{.2cm}
\end{center} 
\begin{center}
{\bf Table 2}
\end{center}

\newpage

\begin{figure}[h]
\centerline{\psfig{figure=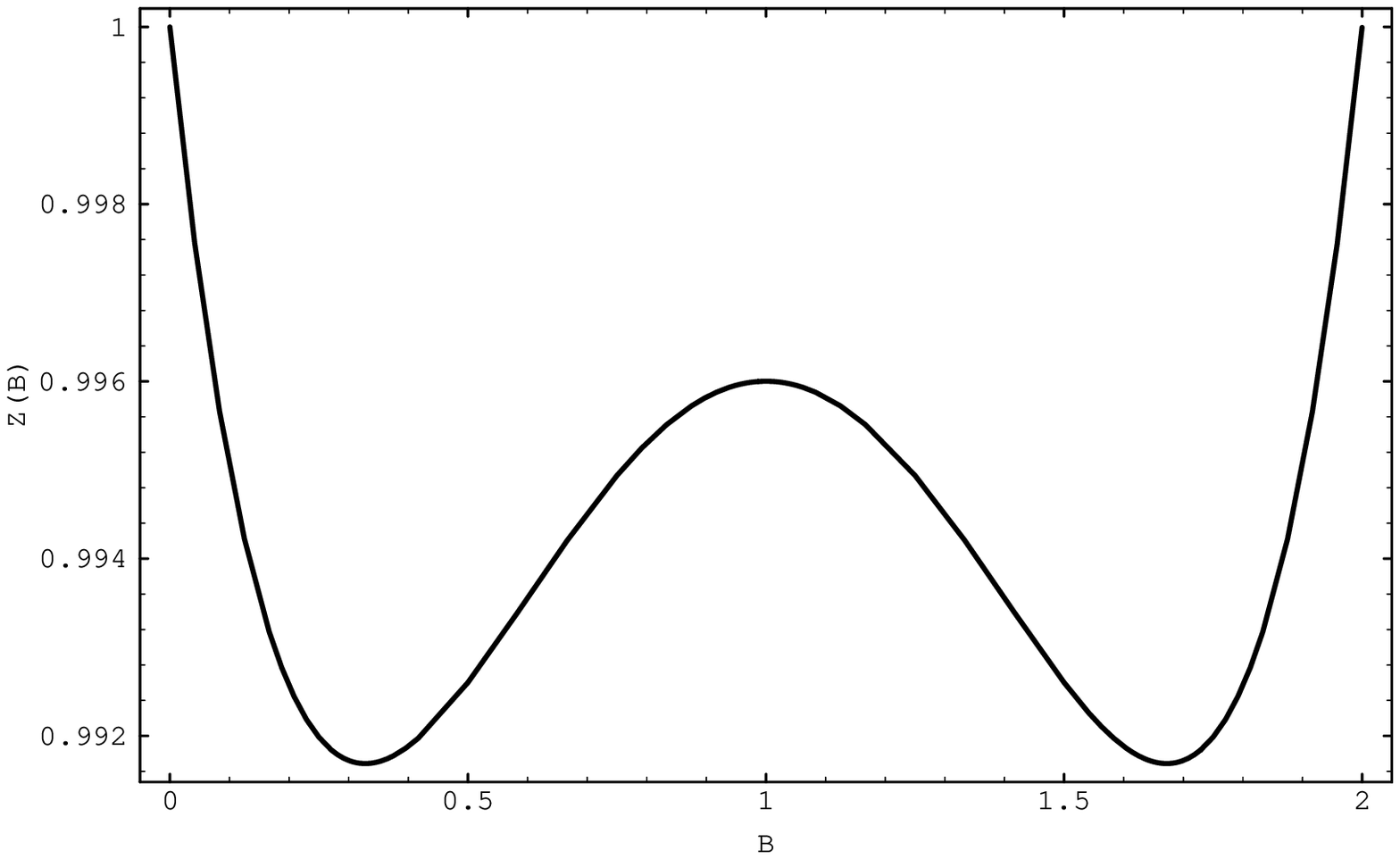,height= 5.5in}}
\begin{center}
{\bf Figure 1:} Plot of the wave function renormalization constant
$Z(B)$ of the Bullough--Dodd model.
\end{center}
\end{figure}

\end{document}